# Are we teaching students to think like scientists?

J. Christopher Moore[*] and Louis J. Rubbo[†]

*Department of Chemistry and Physics*

*Coastal Carolina University, Conway, SC 29528*

**ABSTRACT**

University courses in conceptual physics and astronomy typically serve as the terminal science experience for non-science majors. Significant work has gone into developing research-verified pedagogical methods for the algebra- and calculus-based physics courses typically populated by natural and physical science majors; however, there is significantly less volume in the literature concerning the non-science population. This is quickly changing, and large, repeatable gains on concept tests are being reported. However, we may be losing sight of what is arguably the most important goal of such a course: development of scientific reasoning. Are we teaching this population of students to think like scientists?

[*] email: moorejc@coastal.edu; phone: 843-349-2985
[†] email: lrubbo@coastal.edu; phone: 843-349-6489

University courses in conceptual physics and astronomy typically serve as the terminal science experience for non-science majors. Significant work has gone into developing research-verified pedagogical methods for the algebra- and calculus-based physics courses typically populated by natural and physical science majors; however, there is significantly less volume in the literature concerning the non-science population. This is quickly changing, and large, repeatable gains on concept tests are being reported. However, we may be losing sight of what is arguably the most important goal of such a course: development of scientific reasoning. Are we teaching this population of students to think like scientists?

A recent discussion in *TPT* has led us to examine the central focus of our courses for non-science majors.[1,2] Like Lasry, Finkelstein and Mazur, we certainly do not believe that this population of students are "too dumb" for physics, or that physics is in a "different category" of hard accessible only to certain students such as science majors. However, there are very real differences in the two populations, especially when considering interest level, formal preparation, and prior development of scientific reasoning skills. It is the latter that we focus on in this article, since we believe development of scientific reasoning should be a central goal for these types of courses.

In particular, reasoning and metacognition development[3] are essential if we hope to elevate non-science students to "expert-like" status with respect to problem solving, understanding and applying abstract concepts, and shifting between multiple representations.[4] However, non-science majors enter the classroom with a disadvantage not necessarily shared by their self-selecting science major peers. Non-scientists struggle with basic scientific reasoning patterns, which can hinder their growth in the course.

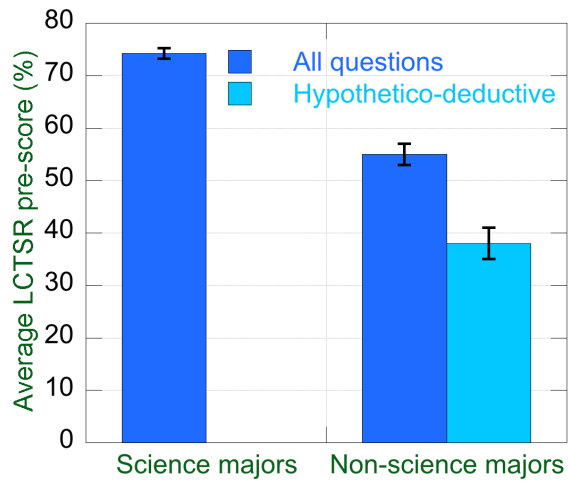

Figure 1: Average scores on the LCTSR before instruction for science (N=1061) and non-science (N=68) majors. Hypothetico-deductive scores where obtained from LCTSR questions 21-24.

We have found that students in our conceptual physics and astronomy courses score significantly lower on Lawson's Classroom Test of Scientific Reasoning (LCTSR) compared to students enrolled in courses typically populated with science majors. The LCTSR assesses reasoning patterns such as proportional reasoning, control of variables, probability reasoning, correlation reasoning and hypothetico-deductive reasoning.[5] Figure 1 shows average LCTSR pre-instruction scores (N = 1061, avg = 74.2%) for freshman science and engineering majors enrolled in a calculus-based introductory physics course, as reported by Bao, et al.[6] The LCTSR was also administered to students taking a conceptual physics or astronomy course with one of the authors during the past three years. As shown in fig. 1, this population of students scores significantly lower (N = 68, avg. = 54%) than their scientist counterparts. Of particular interest, scores on LCTSR questions designed to test application of hypothetico-deductive reasoning, which can arguably be called the "scientific method," average to an abysmal 38%.

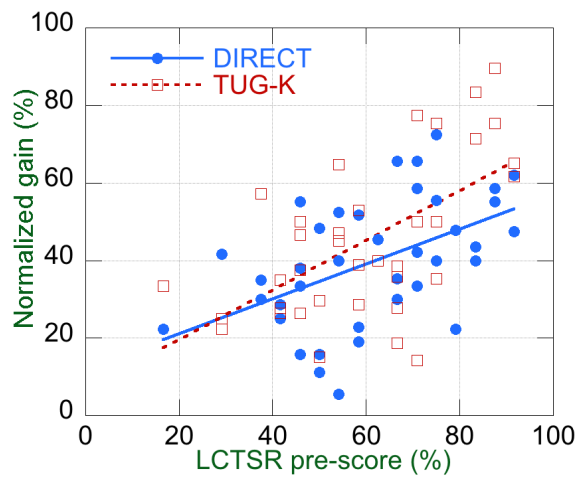

Figure 2: Normalized gain on DIRECT (blue filled circles) and TUG-K (red hollow squares) versus LCTSR pre-instruction scores for non-science majors taking a conceptual physics course. DIRECT: slope=0.45 and $r$=0.50; TUG-K: slope=0.64 and $r$=0.59.

Acknowledgement of this dramatic difference in reasoning ability is important for development of good pedagogy, considering scientific reasoning has been linked to student gains in conceptual knowledge for both non-scientist and scientist populations. Coletta and Phillips observed a strong correlation between normalized gain on the Force Concept Inventory (FCI) and pre-instruction LCTSR scores.[7] As shown in fig. 2, during assessment for our conceptual physics courses over the past two years, we have observed similar strong correlations between pre-instruction LCTSR scores and normalized gain on two concept inventories, the Determining and Interpreting Resistive Electric circuits Concept Test (DIRECT)[8] and the Test for Understanding Graphs -- Kinematics (TUG-K).[9]

Strong correlations are seen for content requiring higher-order and more abstract reasoning. With a slope of linear fit of 0.64 and $r$=0.59, the correlation between TUG-K normalized gain and LCTSR score is similar to that seen for the FCI and stronger than the correlation observed for the DIRECT assessment (slope=0.45 and *r*=0.50). This is not surprising, considering the TUG-K tests a student's ability to move between multiple representations, which rely on higher-order and more abstract thinking.[10] The FCI assesses a student's knowledge and application of the abstract concept of force. Lawson would classify these concepts as hypothetical (motion) and theoretical (force), requiring advanced reasoning development to achieve success.[11] A weaker correlation between DIRECT gains and LCTSR scores could be because strong scores are possible on DIRECT via good observation and retention from well-designed inquiry-based activities; many of the questions are descriptive requiring proficiency only in descriptive level reasoning. This suggests that if we wish to push our non-science students past the lower three levels of Bloom's Taxonomy of Educational Objectives,[12] then we need our courses to focus explicitly on scientific reasoning early and often.

Even with significant disadvantages, substantial gains in content knowledge can still be obtained in conceptual physics and astronomy courses, especially when those courses are designed around a research-verified, active-engagement curriculum. For the conceptual physics and astronomy course, respectively, the authors use a large-enrollment implementation of *Physics by Inquiry* (PbI)[13] and *Lecture-Tutorials for Introductory Astronomy*.[14] Shown in fig. 3 are average normalized learning gains on DIRECT, TUG-K and the Star Properties Concept Inventory (SPCI)[15] for students enrolled in our courses over the past three years. Although lower than reported for students completing some active-engagement algebra- and calculus-based courses, these gains are still significant.

Even though we have been relatively successful with content, we have failed to improve reasoning ability. As seen in fig. 3, average normalized gains on the LCTSR for both physics and astronomy students are essentially equivalent to zero. This is particularly surprising for the conceptual physics course, which via PbI is completely designed around the process of scientific inquiry. Of course, this is not to suggest that gains in reasoning are unachievable. The content-specific education literature in other disciplines suggests that explicit intervention is necessary to improve reasoning.[3,4,16] In fact, we are beginning to see significantly larger gains in scientific reasoning via explicit instruction during our most recent courses, though these observations are preliminary.

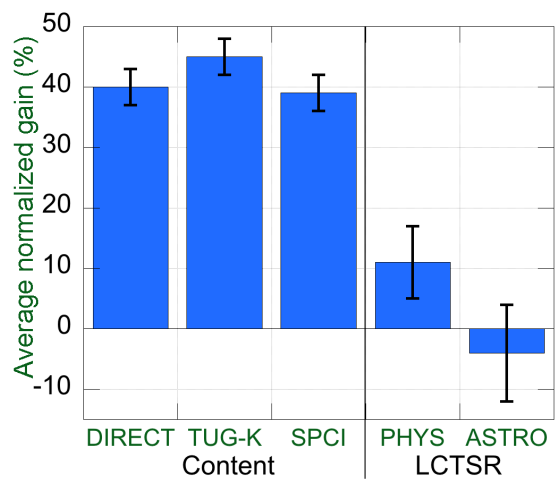

Figure 3: Average normalized gain on DIRECT (N=39), TUG-K (N=40) and SPCI (N=28) for non-science majors taking either a conceptual physics or astronomy course. Average normalized gains on the LCTSR during these courses are also shown.

Development of scientific reasoning is not only a necessary means to an end (making their thinking more scientific so that they can better grasp the content); it is also a justifiable end in and of itself. We should expect our courses to affect our students beyond the classroom. Particularly for non-scientists, a broader approach should be expected since these types of courses are typically their terminal experience in formal science education. Are students in our courses learning to think like scientists? Do we care? Our purpose in writing this paper is to continue the discussion about how we should go about designing our courses for the non-science major. Specifically, is development of scientific reasoning an important goal? If yes, then are we currently achieving that goal? At least in the case of the authors, the answer to that question is no. We are working on that, and we hope others will join us.

**Biographical Sketch**

*Christopher Moore* is an Assistant Professor of Physics at Coastal Carolina University and a former high-school physics teacher. He earned a M.S. in Applied Physics and a Ph.D. in Chemical Physics from Virginia Commonwealth University. He has been involved with developing workshops for K-8 teachers on reasoning in the physical sciences.

*Louis Rubbo* is an Assistant Professor of Astronomy at Coastal Carolina University. He earned a Ph.D. in Physics from Montana State University. He has been involved with developing new educational materials in astronomy and public outreach activities for K-12 students.